\documentclass{emulateapj}
\usepackage{apjfonts}
\usepackage{rotating}

\newcommand{\pivec}{\mbox{\boldmath $\pi$}}

\lefthead{HAN ET AL.} 
\righthead{MICROLENSING MULTIPLE-PLANET SYSTEM}

\begin{document}
\title{
The Second Multiple-Planet System Discovered by Microlensing:
OGLE-2012-BLG-0026L\MakeLowercase{b}, \MakeLowercase{c}, \\
A Pair of Jovian Planets Beyond the Snow Line}

\author{
C. Han$^{1,17,19}$,
A. Udalski$^{2,18}$,
J.-Y. Choi$^{1,17}$,
J. C. Yee$^{5,17}$,
A. Gould$^{5,17}$, 
G. Christie$^{6,18}$, 
T.-G. Tan$^{10,17}$\\
and\\
M. K. Szyma\'nski$^{2}$, 
M. Kubiak$^{2}$, 
I. Soszy\'nski$^{2}$, 
G. Pietrzy\'nski$^{2,3}$, 
R. Poleski$^{2}$, 
K. Ulaczyk$^{2}$, 
P. Pietrukowicz$^{2}$, 
S. Koz{\l}owski$^{2}$, 
J. Skowron$^{5,17}$, 
{\L}. Wyrzykowski$^{2,4}$\\
(The OGLE Collaboration),\\
L. A. Almeida$^{16}$,
V. Batista$^{5}$,
D. L. Depoy$^{7}$, 
Subo Dong$^{8}$, 
J. Drummond$^{13}$, 
B.S. Gaudi$^{5}$, 
K.-H. Hwang$^{1}$, 
F. Jablonski$^{16}$,
Y.-K. Jung$^{1}$, 
C.-U. Lee$^{14}$, 
J.-R. Koo$^{14}$,
J. McCormick$^{9}$, 
L. A. G. Monard$^{11}$, 
T. Natusch$^{6}$,
H. Ngan$^{6}$,
H. Park$^{1}$, 
R. W. Pogge$^{5}$, 
Ian Porritt$^{12}$, 
I.-G. Shin$^{1}$\\
(The $\mu$FUN Collaboration), 
}

\bigskip\bigskip
\affil{$^{1}$Department of Physics, Institute for Astrophysics, Chungbuk National University, Cheongju 371-763, Korea}
\affil{$^{2}$Warsaw University Observatory, Al. Ujazdowskie 4, 00-478 Warszawa, Poland}
\affil{$^{3}$Universidad de Concepci\'{o}n, Departamento de Astronomia, Casilla 160-C, Concepci\'{o}n, Chile}
\affil{$^{4}$Institute of Astronomy, University of Cambridge, Madingley Road, Cambridge CB3 0HA, United Kingdom} 
\affil{$^{5}$Department of Astronomy, Ohio State University, 140 West 18th Avenue, Columbus, OH 43210, USA}
\affil{$^{6}$Auckland Observatory, Auckland, New Zealand}
\affil{$^{7}$Dept. of Physics, Texas A\&M University, College Station, TX, USA}
\affil{$^{8}$Institute for Advanced Study, Einstein Drive, Princeton, NJ 08540, USA}
\affil{$^{9}$Farm Cove Observatory, Centre for Backyard Astrophysics, Pakuranga, Auckland, New Zealand}
\affil{$^{10}$Perth Exoplanet Survey Telescope, Perth, Australia}
\affil{$^{11}$Klein Karoo Observatory, Calitzdorp, and Bronberg Observatory, Pretoria, South Africa}
\affil{$^{12}$Turitea Observatory, Palmerston North, New Zealand}
\affil{$^{13}$Possum Observatory, Patutahi, Gisbourne New Zealand} 
\affil{$^{14}$Korea Astronomy and Space Science Institute, 776 Daedukdae-ro, Yuseong-gu, Daejeon 305-348, Republic of Korea} 
\affil{$^{15}$Institute for Radiophysics and Space Research, AUT University, Auckland, New Zealand}
\affil{$^{16}$Instituto Nacional de Pesquisas Espaciais, S\~{a}o Jos\'{e} dos Campos, SP, Brazil}
\affil{$^{17}$The $\mu$FUN Collaboration}
\affil{$^{18}$The OGLE Collaboration}
\affil{$^{19}$Corresponding author}

\begin{abstract}
We report the discovery of a planetary system from observation of the
high-magnification microlensing event OGLE-2012-BLG-0026.  The lensing 
light curve exhibits a complex central perturbation with multiple features. 
We find that the perturbation was produced by two planets located near 
the Einstein ring of the planet host star. We identify 4 possible solutions 
resulting from the well-known close/wide degeneracy. By measuring both the 
lens parallax and the Einstein radius, we estimate the physical parameters 
of the planetary system. According to the best-fit model, the two planet masses 
are $\sim 0.11\ M_{\rm J}$ and $0.68\ M_{\rm J}$ and they are orbiting a G-type 
main sequence star with a mass $\sim 0.82\ M_\odot$.  The projected separations 
of the individual planets are beyond the snow line in all four solutions, 
being $\sim 3.8$ AU and 4.6 AU in the best-fit solution. The deprojected 
separations are both individually larger and possibly reversed in order.
This is the second multi-planet system with both planets beyond the 
snow line discovered by microlensing. This is the only such a system 
(other than the Solar System) with measured planet masses without $\sin i$
degeneracy.  The planetary system is located at a distance 4.1 kpc from the 
Earth toward the Galactic center.  It is very likely that extra light from 
stars other than the lensed star comes from the lens itself.  If this is 
correct, it will be possible to obtain detailed information about the 
planet-host star from follow-up observation.
\end{abstract}


\section{Introduction}

According to the standard theory of planet formation, rocky planets
form in the inner part of the protoplanetary disk of a star, within 
the snow line, where the temperature is high enough to prevent the 
condensation of water and other substances onto grains \citep{raymond07}. 
This results in coagulation of purely rocky grains and later in the 
formation of rocky planets.  On the other hand, formation of giant 
planets starts at some distance from the host star, beyond the snow 
line, where the low temperature of the protoplanetary disk allows for 
condensation of water ice, enabling more rapid accumulation of solid 
material into a large planetary core. The core accretes surrounding 
gas and later evolves into a gas giant planet \citep{pollack96}.  
Due to the difference in the conditions of formation, rocky terrestrial 
and gas giant planets form at different regions of planetary systems 
\citep{ida04}.

Planet formation theories were initially based entirely on the Solar
System. Planet search surveys using the radial velocity and transit
methods have discovered numerous planets, including many gas
giants. Most of the giant planets have short-period orbits and are
thus much closer to their parent stars than the giant planets in the
Solar System. These discoveries have inspired new theoretical work,
such as various migration theories to resolve the contradiction
between the present-day and assumed-birth locations of giant planets.
However, it is essential to discover more planets located beyond the
snow line to bridge the gap between the Solar System and exo-planet
systems.

\begin{figure*}[ht]
\epsscale{0.8}
\plotone{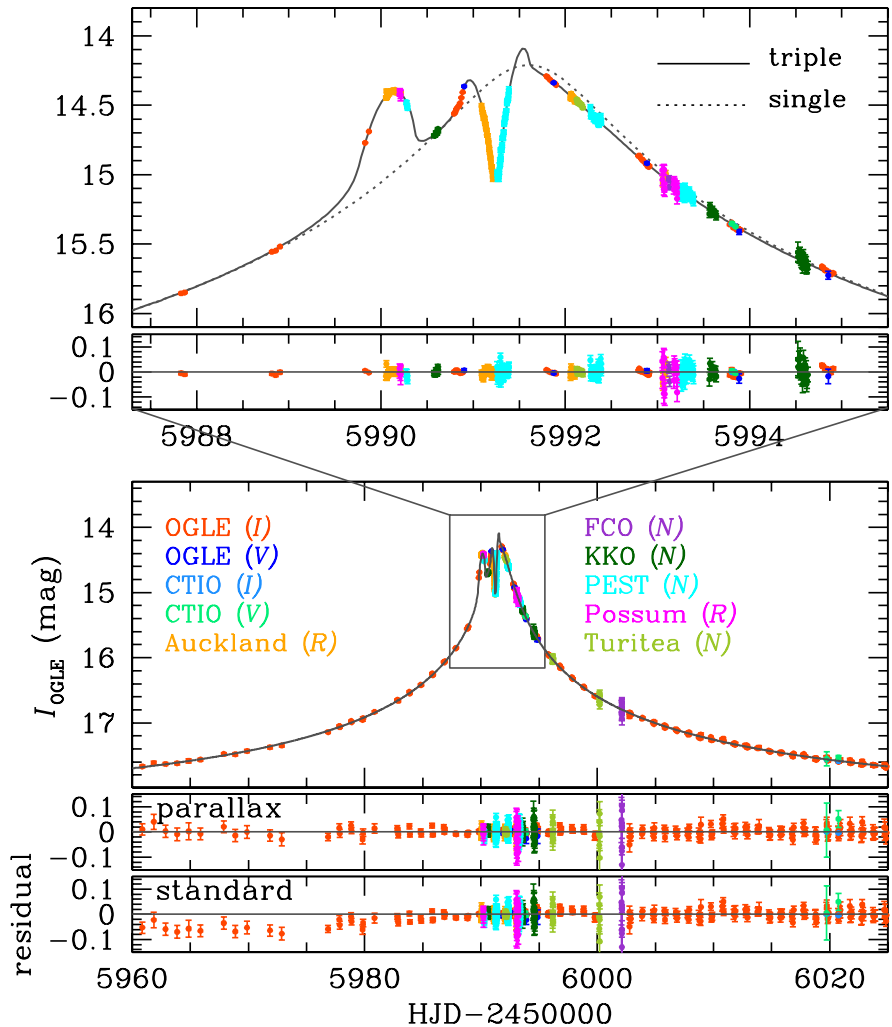}
\caption{\label{fig:one}
Light curve of OGLE-2012-BLG-0026.  The lower panel shows the whole 
view of the light curve and the upper panel shows an enlarged view 
of the perturbation.  Data points taken from different observatories 
are marked by different colors along with the passbands of observation. 
The passband notation ``N'' represents that no filter is used.  The 
dotted and solid curves represent the solutions from single and triple 
lensing modeling, respectively.  The two residuals in the lower panel 
are presented to show the difference between the fits of the models 
with and without the parallax effect.
}\end{figure*}

Microlensing is an important tool that can provide observational
evidence to check planet formation theories.  The planetary signal 
in a microlensing event is a short-term perturbation to the smooth 
standard light curve of the lensing event induced by the planet's 
host star \citep{mao91, gould92b}.  The perturbation is caused by 
the approach of a lensed star (source) close to the planet-induced 
caustics.  The caustic is an optical term representing the envelope 
of concentrated light that forms when light ray is refracted by a 
curved surface, e.g. a curved region of bright light when light 
shines on a drinking glass filled with water or a rippling pattern 
at the bottom of a swimming pool. For gravitational bending of light, 
caustics represents the positions on the source plane at which the 
lensing magnification of a point source becomes infinity due to the 
singularity in the lens-mapping equation. When a caustic is formed 
by a planetary system, it forms multiple sets of closed curves each 
of which is composed of concave curves that meet at cusps. Among them, 
one set is always located close to the planet's host star (central 
caustic) and other sets are located away from the star (planetary 
caustic). Due to these caustic locations, a perturbation induced by 
a central caustic (central perturbation) occurs near the peak of a 
high-magnification event, while a perturbation induced by a planetary 
caustic occurs on the wings of a lensing light curve.  The caustic 
size, and thus the planet detection sensitivity, is maximized when 
the planet is located in the region near the Einstein ring of the 
host star, which is often referred to as the {\it lensing zone}. 
For a typical Galactic lensing event, the physical size of the 
Einstein radius is $r_{\rm E} \sim 3.0 (M/M_\odot)^{1/2}$ AU, where 
$M$ is the mass of the lens.  This is similar to the snow line of 
$2.7(M/M_\odot)^p$ AU, where $0.7\la p \la 2$ \citep{ida05, kennedy08}. 
Therefore, microlensing is an efficient method for detecting cool 
planets located at or beyond the snow line.

Microlensing is also sensitive to multiple-planet systems.
Microlensing detection of multiple planets is possible because all 
planets in the lensing zone affect the magnification pattern around 
the primary lens \citep{griest98}, and thus the signatures of the 
multiple planets can be detected in the light curve of a high 
magnification event \citep{gaudi98}. Indeed, the first two-planet 
microlensing system was discovered through this channel \citep{gaudi08, 
bennett10}.

In this paper, we report the second discovery of a multiple-planetary 
system composed of two planets located beyond the snow line discovered 
by using the microlensing method.

\section{Observation}

OGLE-2012-BLG-0026 was discovered by the Optical Gravitational
Lensing Experiment Early Warning System (OGLE EWS; Udalski 2003) 
on 2012-Feb-13 toward the Galactic bulge at $(\alpha,\delta)_{\rm J2000}
=(17^{\rm h}34^{\rm m}18^{\rm s}\hskip-2pt.70, -27^{\circ}08'33''\hskip-2pt.9)$,
i.e., $(l,b)=(0.195^\circ,3.066^\circ)$, based on $I$-band observations
with the 1.3m Warsaw telescope at Las Companas Observatory in Chile.
At HJD' (= HJD-2450000) = 5989.3 (Mar 2), the Microlensing Follow-Up
Network ($\mu$FUN) issued an alert, saying that the event would reach 
high magnification ($A>100$) 2.5 days hence, and urging immediate 
observations.   Given this extremely early date (3.5 months before 
the target reaches opposition), and hence very short observing window 
from any individual site, this quick alert was crucial to mobilizing 
near-24-hour coverage from multiple sights.  $\mu$FUN covered the peak 
with observations from 6 telescopes: the 0.4 m at Auckland Observatory, 
0.4 m at Farm Cove Observatory (FCO), 0.4 m Possum Observatory, 0.4 m 
Turitea Observatory in New Zealand, 0.4 m Klein Karoo Observatory (KKO) 
in South Africa, and 0.3 m Perth Exoplanet Survey Telescope (PEST) in 
Australia.  The Auckland data were taken in $R$ band, and the remainder 
were unfiltered.  In addition, $\mu$FUN obtained supplementary $V/I/H$ 
color observations using the 1.3 m SMARTS telescope at Cerro Tololo 
Inter-American Observatory (CTIO).  Follow-up observation was stopped 
after the perturbation, but normal monitoring of the event was continued 
by the OGLE survey until the event returned to its baseline brightness, 
and beyond.

Photometric reductions  of data were carried out using photometry codes 
developed by the individual groups. The OGLE data were reduced by a
photometry pipeline that applies the image subtraction method based 
on the Difference Image Analysis technique \citep{alard98} and further 
developed by \citet{wozniak00} and \citet{udalski03}. Initial reduction 
of the $\mu$FUN data were processed using a pipeline based on the DoPHOT 
software \citep{schechter93}.  The data were reprocessed by using the 
pySIS package \citep{albrow09} to refine the photometry.

Figure \ref{fig:one} shows the light curve of OGLE-2012-BLG-0026.  The 
magnification of the lensed source star flux reached $A_{\rm max}\sim 129$ 
at the peak.  The perturbation is localized near the peak region of the 
light curve. It exhibits a complex structure with multiple features.  At 
a glance, the perturbation is composed of two parts. One part, at 
${\rm HJD}'\sim 5990.2$ has a positive deviation with respect to the 
unperturbed single-lens light curve.  The other part, centered at 
${\rm HJD}'\sim 5991.2$ has a negative deviation.  Such a complex 
perturbation pattern is unusual for central perturbations.

\section{Interpretation of light curve}

To interpret the observed light curve, we initially conducted modeling 
based on a standard two-mass lens model. Description of a binary-lens 
light curve requires 3 single lensing parameters plus another 3 parameters
related to the binary nature of the lens. The single lensing parameters
include the moment of the closest lens-source approach, $t_0$, the projected
lens-source separation at that moment, $u_0$, and the Einstein time scale,
$t_{\rm E}$, of the event. The Einstein time scale is defined as the time
required for the source to cross the angular Einstein radius of the lens,
$\theta_{\rm E}$. The binary lensing parameters are the projected binary
separation in units of the Einstein radius, $s$, the mass ratio between
the lens components, $q$, and the angle between the source trajectory and
the binary axis, $\alpha$. Since central perturbations usually involve 
caustic crossings or approaches, an additional parameter of the source 
radius in units of $\theta_{\rm E}$, $\rho_\star$ (normalized source radius), 
is needed to account for the attenuation of lensing magnifications by the 
finite size of the source. A central perturbation in the light curve of a 
high magnification event can be produced either by a planetary companion 
positioned in the lensing zone or by a binary companion located away from 
the Einstein radius \citep{han08, han09}.  From a thorough search considering 
both possible cases of central perturbations, however, we could not find any 
solution that explains the observed perturbation.

We therefore consider the possibility of an additional lens companion. 
\citet{han05} pointed out that many cases of central perturbations induced 
by multiple planets can be approximated by the superposition of the 
single-planet perturbations in which the individual planet-primary pairs 
act as independent binary lens systems. Based on this ``binary superposition'' 
approximation, we then conduct another binary lens modeling of the two light 
curves each of which is prepared by including only a single perturbed region. 
The first light curve includes the negative perturbation region during 
$5990.5 < {\rm HJD}' < 5992.0$ and the second includes the positive 
perturbation region during $5989.0 < {\rm HJD}' < 5990.5$. From this, we 
find that the two partial light curves are well fitted by binary lensing 
light curves with two different planetary companions.

Based on the parameters of the individual planetary companions, we then
conduct triple-lens modeling, which requires to include three additional 
parameters related to the second companion. These are the normalized 
projected separation from the primary, $s_2$, the mass ratio between the 
second companion and the primary, $q_2$, and the position angle of the 
second planet with respect to the line connecting the primary and the 
first planet, $\psi$. We denote the parameters related to the first planet 
as $s_1$ and $q_1$. We note that the first and second planets are designated 
as the ones producing the negative and positive perturbations, respectively. 
With the initial values of the triple-lens parameters obtained from the binary 
superposition approximation, we locate the $\chi^2$ minimum using the Markov 
Chain Monte Carlo (MCMC) method.

\begin{figure}[ht]
\epsscale{1.15}
\plotone{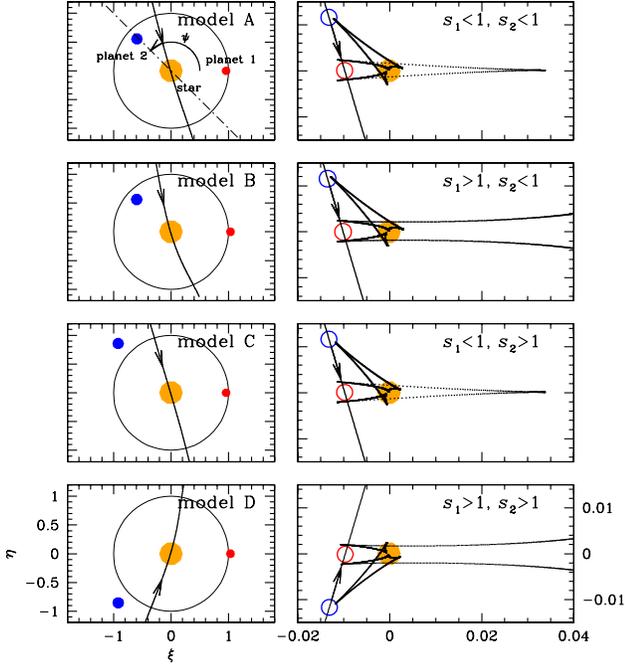}
\caption{\label{fig:two}
Four possible geometries of the lens systems. Each of the left panels
shows the relative positions of the two planets with respect to the host
star. Coordinates are centered at the position of the host star and all
lengths are normalized by the Einstein radius. The circle with a radius
1.0 represents the Einstein ring. Also marked is the trajectory of the
source star (a curve with an arrow). Right panels show an enlarged view
of the central region around the host star. The cuspy feature represents
the caustics. The two small empty dots represent the source star at the 
moments of the major perturbations. The size of the dots is scaled to 
represent the source size.
}\end{figure}

We find that consideration of the parallax effect is needed to precisely 
describe the light curve. The parallax effect is caused by the change of 
the observer's position due to  the orbital motion of the Earth around 
the Sun during the event \citep{gould92a}.  For OGLE-2012-BLG-0026, the 
event time scale is of order 90 days, which comprises a significant 
fraction of the Earth's orbital period, and thus parallax can affect
the lensing light curve. We therefore introduce two additional parallax 
parameters of $\pi_{{\rm E},N}$ and $\pi_{{\rm E},E}$, which represent 
the two components of the lens parallax vector $\pivec_{\rm E}$ projected 
on the sky in the north and east equatorial coordinates, respectively. 
The direction of the parallax vector corresponds to the lens-source relative 
motion in the frame of the Earth.  The size of the parallax vector is defined 
as the ratio of Earth's orbit to the physical Einstein radius projected on 
the observer plane, i.e. $\pi_{\rm E}={\rm AU}/[\theta_{\rm E}/(D_{\rm L}^{-1}-
D_{\rm S}^{-1})]$, where $D_{\rm L}$ and $D_{\rm S}$ are the distances 
to the lens and source star, respectively.  We find that the parallax 
effect improves the fit by $\Delta\chi^2\sim 240$.  We note that 
``terrestrial'' parallax effect due to non-cospatial observatories on 
the Earth \citep{gould09} is not detected since the separation between 
the 2 observatories covering the caustic crossings (Auckland and PEST)  
is not big enough.

For high-magnification events, planets described by $(s,q)$ and
$(s^{-1},q)$ induce nearly identical caustics and thus nearly identical 
light curves \citep{griest98}.  This is known as the close/wide degeneracy 
in lensing light curves.  We, therefore, investigate other solutions 
with initial values of the projected separations based on the first solution. 
From this search, we find 4 sets of solutions with ($s_1<1$, $s_2<1$), 
($s_1>1$, $s_2<1$), ($s_1<1$, $s_2>1$), and ($s_1>1$, $s_2>1$). We denote 
the individual models as ``A'', ``B'', ``C'', and ``D''.

If the source were exactly on the ecliptic, it would exactly obey the
``ecliptic degeneracy'', which takes $(u_0,\alpha,\pi_{{\rm E},\perp})
\rightarrow -(u_0,\alpha,\pi_{{\rm E},\perp})$ \citep{skowron11}.
Here $\pi_{{\rm E},\perp}$ is the component of $\pivec_{\rm E}$ perpendicular
to the ecliptic at the projected position of the Sun at $t_0$, which in the 
present case is almost exactly north. OGLE-2012-BLG-0026 lies $<4^\circ$ 
[$(\lambda,\beta)_{J2000}=(264.28^\circ,-3.83^\circ)$] from the ecliptic. 
Hence it is not surprising that we find almost perfect ecliptic degeneracy 
for all four solutions, with $\Delta\chi^2 \lesssim 1$.  We do not show 
these redundant solutions.

In principle, one can detect lens orbital motion, and even if not, tangential 
orbital motion can be degenerate with $\pi_{{\rm E},\perp}$ \citep{batista11},
which would degrade the parallax measurement. We fit for all possible 
orientations of co-planar circular orbital motion (hence two additional 
parameters to define the orbit plane). However, we find, first that there 
is no improvement in the fit ($\Delta\chi^2 \lesssim 1$), and second that there is 
no degeneracy with $\pi_{{\rm E},\perp}$. We therefore do not further 
consider orbital motion.

In Table \ref{table:one}, we present the four solutions. In Figure \ref{fig:two}, 
we also present the geometries of the lens system corresponding to the individual 
solutions.  We find that both planets responsible for the perturbation are located 
near the Einstein ring of the host star. The negative perturbation was produced 
by the passage of the source trajectory on the backside of the central caustic 
induced by a planet located almost on the Einstein ring of the host star, while 
the positive perturbation was produced when the source passed the sharp front 
tip of the arrowhead-shaped central caustic induced by another planet located 
at a position slightly away from the Einstein ring.  The position angle 
of the second planet with respect to the first planet is $\psi\sim 220^\circ$.  
We note that although the close/wide degeneracy causes ambiguity in the 
star-planet separation, the mass ratios of the two degenerate solutions 
are similar to each other. We find that the mass ratios are $q_1 \sim 1.3 
\times 10^{-4}$ and $q_2 \sim 7.8 \times 10^{-4}$.

\section{Physical Parameters}

The combined measurements of the angular Einstein radius and the lens parallax 
enable an estimate of the lens mass and distance.  With these values, the 
mass and distance are determined by $M=\theta_{\rm E}/(\kappa\pi_{\rm E})$ 
and the $D_{\rm L}={\rm AU}/(\pi_{\rm E}\theta_{\rm E}+\pi_{\rm S})$, where 
$\kappa= 4G/(c^2{\rm AU})$ and $\pi_{\rm S}={\rm AU}/D_{\rm S}$ is the parallax 
of the source star \citep{gould00}.  By modeling the Galactic bulge as a bar, 
the distance to the source star is estimated by $D_{\rm S}=D_{\rm GC}(\cos\ell 
+ \sin\ell\cos\phi/\sin\phi)$, where $D_{\rm GC}=8.16$ pc is the Galactocentric 
distance \citep{nataf12}, $\ell=0.195^\circ$ is the Galactic longitude of the 
source star, $\phi=40^\circ$ is the bar orientation angle with respect to the 
line of sight.  This results in $D_{\rm S}=8.13$ kpc, corresponding to 
$\pi_{\rm S}=0.123$ mas.

\begin{figure}[ht]
\epsscale{1.15}
\plotone{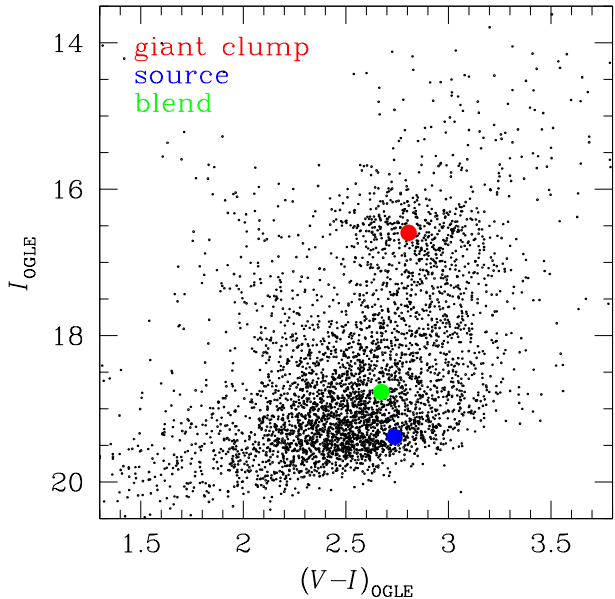}
\caption{\label{fig:three}
Locations of the source star and blend in the color-magnitude diagram.
Also marked is the position of the giant clump centroid that is used 
to calibrate the color and magnitude of the lensed star.  Black dots 
are stars in the field where the source star is located.
}\end{figure}

The Einstein radius is given by $\theta_{\rm E}=\theta_\star/\rho_\star$
where $\theta_\star$ is the angular radius of the source, which we evaluate 
from its de-reddened color $(V-I)_0$ and brightness $I_0$ using an instrumental 
color-magnitude diagram \citep{yoo04}.  We first measure the offset 
$\Delta[(V-I),I] = (-0.07,2.79)$ mag of the source from the clump, whose de-reddened 
position is known independently, $[(V-I),I]_{0,\rm c} = (1.06,14.45)$ mag
\citep{bensby11, nataf12}.  This yields the de-reddened color and magnitude 
of the source star $[(V-I),I]_{0,{\rm S}} = (0.99,17.23)$ mag.  Thus the source 
is a bulge subgiant.  Note that \citet{bensby13} estimate a very similar 
color $(V-I)_0=1.00$ mag based on a high-resolution VLT spectrum taken at high 
magnification.  Figure \ref{fig:three} shows the locations of the source 
star and the centroid of the giant clump in the color-magnitude diagram 
that is constructed based on the OGLE data.  Once the de-reddened $V-I$ 
color of the source star is measured, it is translated into $V-K$ color 
by using the $V-I$ versus $V-K$ relations of \citet{bessell88} and then 
the angular source radius is estimated by using the relation between the 
$V-K$ color and the angular radius given by \citet{kervella04}.  The measured 
angular source radius is $\theta_\star =1.55 \pm 0.13\ {\mu}{\rm as}$. Then, 
the Einstein radius is estimated as $\theta_{\rm E}=0.91 \pm 0.09$ mas.

In Table \ref{table:two}, we list the physical parameters of the 
planetary system determined based on model D, which provides the 
best fit. Other models are similar.  According to the estimated mass 
of $\sim 0.82\ M_\odot$, the host star of the planets is a G-type main 
sequence. The individual planets have masses of $\sim 0.11\ M_{\rm J}$ 
and $0.68\ M_{\rm J}$.  Their projected separations are $\sim$ 3.8 AU 
and 4.6 AU, respectively, and thus both planets are located well beyond 
the snow line 1.8 -- 2.4 AU of the host star. The planetary system is 
located at a distance $\sim$ 4.1 kpc from the Earth toward the Galactic 
bulge direction.


To be noted is that the blend is likely to be the lens itself.  Blended 
light refers to the extra light coming from stars other than the lensed 
star. The source of blended light can be unresolved stars located close 
to the lensed star or the lens itself.  The fraction of blended light 
among total observed light is measured from modeling of the light curve. 
Considering 
the distance to the lens $\sim 4.1$ kpc, one can assume that the lens and 
the source experience roughly the same amount of reddening and extinction. 
Under this approximation, the estimated de-reddened color and brightness 
of the blend determined with the reference of the giant clump are 
$(V-I,I)_{0,{\rm b}}=(0.93,16.62)$ mag. Then, under the assumption that 
the blend is the lens combined with the distance modulus $\sim$ 13.05 mag 
to the lens, the estimated absolute magnitude of the blend is $M_{I,{\rm b}}
\sim 3.6$ mag. Then, the color and the absolute magnitude of the blend roughly 
agree with the color $(V-I)_{0}\sim0.9$ mag and the absolute magnitude 
$M_I\sim 4.2$ mag of a G5V star, that corresponds to the lens, suggesting that 
blended light comes from the host star.  Moreover, from difference imaging 
the magnified and unmagnified images, we find that the blend and source 
have identical positions within measurement error ($\sim 50$ mas), further 
tightening the identification of the blend with the lens.

This identification opens many possibilities for future follow-up observations.
AO imaging could confirm and refine the host star flux measurement by
isolating it from fainter ambient stars. At $I=19$ mag, the host is bright 
enough to obtain a spectroscopic metallicity measurement, which would be 
the first for a microlensing planet host. Finally, late-time astrometry 
could improve the mass and distance determinations by refining the 
$\theta_{\rm E}$ measurement from the magnitude of the proper-motion 
vector \citep{alcock01} and refining the $\pivec_{\rm E}$ measurement from 
its direction \citep{ghosh04}.

\begin{deluxetable}{ll}
\tablecaption{Physical Parameters\label{table:two}}
\tablewidth{0pt}
\tablehead{
\multicolumn{1}{c}{parameter} &
\multicolumn{1}{c}{quantity}
}
\startdata
mass of the host star                        & 0.82 $\pm$ 0.13 $M_\odot$  \\
mass of the first planet                     & 0.11 $\pm$ 0.02 $M_{\rm J}$ \\
mass of the second planet                    & 0.68 $\pm$ 0.10 $M_{\rm J}$ \\
projected separation to the first planet     & 3.82 $\pm$ 0.30 AU \\
projected separation to the second planet    & 4.63 $\pm$ 0.37 AU  \\
distance to the planetary system             & 4.08 $\pm$ 0.30 kpc
\enddata  
\end{deluxetable}


%

\begin{deluxetable*}{lrrrr}
\tablecaption{Lensing Parameters\label{table:one}}
\tablewidth{0pt}
\tablehead{
\multicolumn{1}{l}{parameters} &
\multicolumn{1}{c}{model A} &
\multicolumn{1}{c}{model B} &
\multicolumn{1}{c}{model C} &
\multicolumn{1}{c}{model D} 
}
\startdata
$\chi^2/{\rm dof}$        &  2689.7/2667            &  2681.0/2667            &  2680.8/2667            &  2674.4/2667          \\
$t_0$ (HJD')              &  5991.52 $\pm$ 0.01     &  5991.52 $\pm$ 0.01     &  5991.52 $\pm$ 0.01     &  5991.52 $\pm$ 0.01   \\
$u_0$                     & -0.0094  $\pm$ 0.0001   & -0.0097  $\pm$ 0.0001   & -0.0094  $\pm$ 0.0001   &  0.0092  $\pm$ 0.0001 \\
$t_{\rm E}$ (days)        &  92.45   $\pm$ 1.08     &  89.21   $\pm$ 0.17     &  92.85   $\pm$ 1.11     &  93.92   $\pm$ 0.58   \\
$s_1$                     &  0.957   $\pm$ 0.001    &  1.034   $\pm$ 0.001    &  0.957   $\pm$ 0.001    &  1.034   $\pm$ 0.001  \\
$q_1$ ($10^{-4}$)         &  1.32    $\pm$ 0.02     &  1.37    $\pm$ 0.01     &  1.30    $\pm$ 0.02     &  1.30    $\pm$ 0.01   \\
$s_2$                     &  0.812   $\pm$ 0.004    &  0.819   $\pm$ 0.004    &  1.257   $\pm$ 0.006    &  1.254   $\pm$ 0.006  \\
$q_2$ ($10^{-4}$)         &  7.97    $\pm$ 0.26     &  7.84    $\pm$ 0.25     &  8.04    $\pm$ 0.25     &  7.84    $\pm$ 0.21   \\
$\alpha$                  &  1.283   $\pm$ 0.001    &  1.284   $\pm$ 0.001    &  1.284   $\pm$ 0.001    &  4.999   $\pm$ 0.001  \\
$\psi$                    &  2.389   $\pm$ 0.002    &  2.387   $\pm$ 0.001    &  2.391   $\pm$ 0.001    &  3.891   $\pm$ 0.001  \\
$\rho_\star$ ($10^{-3}$)  &  1.73    $\pm$ 0.02     &  1.81    $\pm$ 0.01     &  1.72    $\pm$ 0.02     &  1.72    $\pm$ 0.01   \\
$\pi_{{\rm E},N}$         & -0.029   $\pm$ 0.054    & -0.097   $\pm$ 0.012    &  0.001   $\pm$ 0.028    & -0.072   $\pm$ 0.051  \\
$\pi_{{\rm E},E}$         &  0.123   $\pm$ 0.009    &  0.137   $\pm$ 0.003    &  0.123   $\pm$ 0.005    &  0.114   $\pm$ 0.004  
\enddata                             
\tablecomments{ 
HJD'=HJD-2450000.
}
\end{deluxetable*}

\acknowledgments 
Work by CH was supported by Creative Research Initiative Program 
(2009-0081561) of National Research Foundation of Korea.
 The OGLE project has received funding from the European Research
Council under the European Community's Seventh Framework Programme
(FP7/2007-2013) / ERC grant agreement no. 246678.
A. Gould and B.S. Gaudi acknowledge support from NSF AST-1103471. 
B.S.  Gaudi, A. Gould, and R.W. Pogge acknowledge support from NASA 
grant NNX12AB99G. Work by J.C. Yee is supported by a National Science
Foundation Graduate Research Fellowship under Grant No. 2009068160. 
S.  Dong's research was performed under contract with the California
Institute of Technology (Caltech) funded by NASA through the Sagan
Fellowship Program.

\end{document}